\documentclass[10pt]{elsart}
\usepackage{graphicx}
\usepackage{subfigure}
\usepackage{latexsym}

\begin{document}
\newcommand{\mrm}{\mathrm}
\newcommand{\n}{{\em n}$^{+}$}
\newcommand{\p}{{\em p}$^{+}$}
\newcommand{\pn} {{\em pn}}
\newcommand{\pit}{{\em p}}
\newcommand{\Pit}{{\em P}}
\newcommand{\nit}{{\em n}}
\newcommand{\Nit}{{\em N}}
\newcommand{\Neq}{\mrm{n}_{\mrm{eq}}/\mrm{cm}^{2}}
\newcommand{\fns}{\footnotesize}
\newcommand{\scrs}{\scriptsize}

%%upright Greek letters (example below: upright "mu")
\newcommand{\greeksym}[1]{{\usefont{U}{psy}{m}{n}#1}}
\newcommand{\umu}{\mbox{\greeksym{m}}}
\newcommand{\udelta}{\mbox{\greeksym{d}}}
\newcommand{\uDelta}{\mbox{\greeksym{D}}}
\newcommand{\uOmega}{\mbox{\greeksym{W}}}
\newcommand{\uPi}{\mbox{\greeksym{P}}}
\newcommand{\ualpha}{\mbox{\greeksym{a}}}

\sloppy
\begin{frontmatter}
\title{Fluence Dependence of Charge Collection of irradiated Pixel Sensors}
\author[psi]{T.\,Rohe\thanksref{corr}}
\author[purdue]{D.\,Bortoletto} \author[unizh]{V.\,Chiochia}
\author[mi]{L. M. Cremaldi}
\author[unibs]{S.\,Cucciarelli} \author[unizh,psi]{A.\,Dorokhov}
\author[psi,unizh]{C.\,Hoermann}
\author[jhu]{D.\, Kim} \author[unibs]{M.\,Konecki}
\author[psi]{D.\,Kotlinski} \author[unizh,psi]{K.\,Prokofiev}
\author[unizh]{C.\,Regenfus}
\author[mi]{D. A. Sanders}
\author[purdue]{S.\,Son} \author[unizh]{T.\,Speer}
\author[jhu]{M.\,Swartz}
\address[psi]{Paul Scherrer Institut, 5232 Villigen PSI, Switzerland}
\address[purdue]{Purdue University - Task G, West Lafayette, IN 47907, USA}
\address[unizh]{Physik Institut der Universit\"at Z\"urich-Irchel,
	8057 Z\"urich, Switzerland}
\address[mi]{Department of Physics and Astronomy,University of Mississippi,
	University, MS 38677, USA}
\address[unibs]{Institut f\"ur Physik
        der Universit\"at Basel, 4056 Basel ,Switzerland}
%\address[psi]{Paul Scherrer Institut, 5232 Villigen PSI, Switzerland}
\address[jhu]{Johns Hopkins University, Baltimore, MD 21218, USA.}

\thanks[corr]{Corresponding author; e-mail: Tilman.Rohe@cern.ch}

\begin{abstract}
The barrel region of the CMS pixel detector will be equipped
with ``n-in-n'' type silicon sensors. They are processed on 
diffusion oxygenated float zone (DOFZ) material, use the moderated
p-spray technique for inter pixel isolation and feature a  bias grid. 
The latter leads to a small fraction of the pixel area to be less sensitive
to particles. In order to quantify this inefficiency
prototype pixel sensors irradiated to particle fluences
between $4.7\times 10^{13}$ and $2.6\times 10^{15}\,\Neq$
have been bump bonded to un-irradiated readout chips and tested
using high energy pions at the H2 beam line of the CERN SPS.
The readout chip allows a non zero suppressed analogue
readout and is therefore well suited to measure the charge
collection properties of the sensors.

In this paper we discuss the fluence dependence of the collected
signal and the particle detection efficiency. Further the position
dependence of the efficiency is investigated.

{\em PACS:} 07.89; 29.40.Gx; 29.40.Wk; 61.80.-x

{\em Key words:} LHC, CMS, tracking, pixel, silicon, radiation hardness.
\end{abstract}
\end{frontmatter}

\section{Introduction}
The CMS experiment, currently under construction
at the Large Hadron Collider (LHC) at CERN (Geneva, Switzerland),
will contain a hybrid pixel detector for tracking and vertexing \cite{cms-pixel}.
In its final configuration it will consist of three barrel layers and two end
disks at each side.

To improve the spatial resolution analogue interpolation between neighboring
channels will be performed. 
%\marginpar{\tiny{Note to referee: B-field was reason for pitch in $r-\phi$
%not vice versa}}
The strong Lorentz deflection in the radial direction
caused by CMS' $4\,{\mrm T}$ magnetic field distributes the signal over
two and more pixels. For this reason the pixel size of $100\times 150\,\umu\mrm{m}^{2}$
was chosen. In the disks, where the charge carrier drift
is minimally affected by the magnetic field, the modules are tilted by
about $20^{\circ}$ with respect to the plane orthogonal to the beam line
to induce charge sharing between pixels.

Because of the harsh radiation environment at the LHC, the technical realization
of the pixel detector is very challenging.
The innermost barrel layer will
be exposed to a fluence of about $3\times 10^{14}\,\Neq$ per year at the full
LHC-luminosity, the second and third layer to about $1.2\times
10^{14}\,\Neq$ and \ $0.6\times 10^{14}\,\Neq$, respectively.
All components of the pixel detector are
specified to remain operational up to a particle fluence of at least
$6\times 10^{14}\,\Neq$.
This implies that parts of the detector will have to be replaced
during the lifetime of the experiment. In the case of a possible luminosity
upgrade of the LHC the particle fluences will be much higher.
For this reason it is necessary to test if the detectors can be operated
at fluences above the ones specified.
The life time of the sensor is limited by insufficient charge collection
caused by trapping and incomplete depletion. As both effects
can be reduced by increasing the sensor bias,  the sensor
design must allow the operation at high bias voltages without
electrical breakdown. For the CMS pixel detector a maximum value of
500-600\,V is foreseen.

In addition to the radiation-induced bulk effects the charge collection properties
of the sensor are also influenced by the pixel design (e.g.
the implant geometry). Therefore, the design has to be optimized to minimize possible
regions with reduced signal collection. The aim of this study is to
investigate the fluence and position dependence of the charge collection
properties in the CMS prototype sensors.

\section{The CMS Pixel Barrel Sensors}

For the sensors of the pixel detector the ``\nit -in-\nit'' concept has
been chosen. Electron collection has the advantage that after irradiation
induced space charge sign inversion, the highest electric field is
located close to the collecting electrodes. %\cite{morris}.
In addition double-sided processing of these devices allows the implementation of
guard rings only on the \pit -side of the sensor, keeping all sensor 
edges at ground potential. The design of the guard rings has been optimized 
in the past \cite{rolf}. The breakdown voltage safely exceeds the
required value of 600\,V.

\begin{figure}
\centering
\includegraphics[width=.48\linewidth]{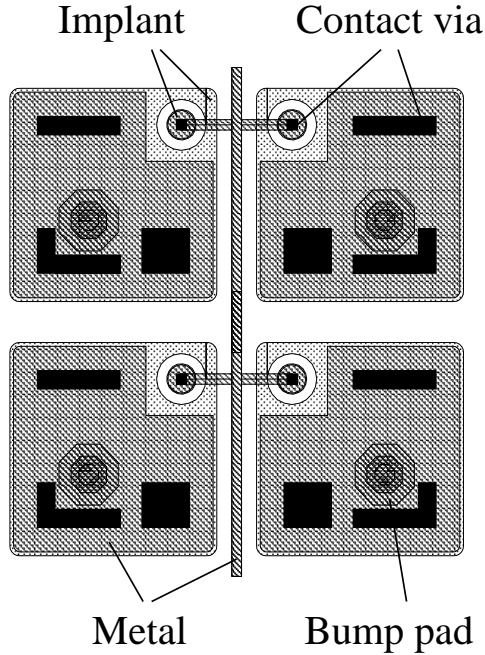}
\caption{Mask layout of the pixel sensors under study.}
        \label{designs01}
\end{figure}

Due to the superior performance after
irradiation and the possibility to implement a bias grid the
moderated  \pit -spray technique was chosen for inter pixel isolation in 
the pixel barrel \cite{ieee03}.
The pixel layout is shown in Fig.~\ref{designs01} and is characterized by small gaps
of $20\,\umu$m between the \n -implants and by a biasing structure
implementing small punch through ``bias dots'' \cite{pixpt}.
They allow on wafer current-voltage (IV) measurements and keep accidentally 
unconnected pixel cells close to ground potential. 
Following the recommendation of the ROSE
collaboration \cite{rose}, oxygen enriched silicon was used to
improve the post irradiation behavior. The thickness of the sensors was
$285\pm 15\,\umu$m.

The pixel size of the sensors investigated in this study was % chosen to
$125\times 125\,\umu\mrm{m}^{2}$ in order to match the readout chip.
Although these dimensions differ from the ones foreseen in CMS we
are confident that the basic charge collection properties presented in this
paper are not affected by the cell size. Other properties, e.g. the spatial
resolution, have to be measured with the final configuration.

\section{Testing Procedure}

In a pixelated device the important parameters
for the performance of a single channel, like pixel
capacitance and leakage current, are independent of the array dimensions.
Therefore the use of miniature sensors does not restrict the validity
of the results. The results presented in this paper were obtained
with sensors containing $22\times 32$ pixels.

After the deposition of the under bump metalization and the indium bumps
the sensors were diced.
Some of them were irradiated at the CERN PS with 24\,GeV
protons\footnote{hardness factor 0.62 \cite{rose}} to
fluences from 0.47 to $26\times 10^{14}\,\Neq$ (see tab.~\ref{tab:par}).
The irradiation was performed without cooling and bias.

In order to avoid reverse annealing the sensors were stored at $-20^{\circ}\,$C after
irradiation and warmed up only for transport and bump bonding. Some of the
samples were annealed for three days at $30^{\circ}$ close to the minimum of
the full depletion voltage \cite{michael_excel}. To sort out defective sensors
all of them were characterized with IV-measurements before and after irradiation.

Miniature sensors were bump bonded to PSI30/AC30\footnote{PSI30 DMILL pixel readout chip
was designed in 1997 at Paul Scherrer Institut, Villigen, Switzerland,
and translated in 1998 to the Honeywell RICMOS~IV process at 1.~Physikalisches
Institut of the RWTH Aachen, Germany.} readout
chips described in detail in~\cite{david}.
This chip was chosen instead of the final CMS-pixel readout chip
because it allows 
a sequential readout of all pixel cells without zero suppression.
The sampling time at the shaper was defined by an external hold signal
provided by a pin-diode with a delay of about 60\,ns.
The peaking times of the preamplifier and the shaper were adjusted
to about 40\,ns by tuning the feedback resistors of the charge sensitive
amplifiers. This setting prevents saturation of the preamplifier and shaper 
up to signals corresponding to about 1.5 minimal ionizing particles (m.i.p.)
but leads to a higher noise.
As the readout chip is not sufficiently radiation hard, 
irradiated sensors were bump bonded to un-irradiated readout chips
Therefore a special bump bonding
procedure without heat application was used.

The bump bonded samples were tested at the CERN-SPS H2 beam line using
$150\,$GeV pions in 2003 and 2004. The pixel device under test was situated
in-between a four layer silicon strip telescope \cite{bt} with an intrinsic 
spatial resolution of about $1\,\umu$m. 
The sensors were cooled to a temperature of $-10^{\circ}\,$C
by water cooled Peltier elements.
The whole set-up was placed in a 3\,T magnet with the $\vec{B}$~field parallel
to the beam (2003) or perpendicular (2004). The pixel detector was set either
normal to the beam ($90^{\circ}$), tilted by a small angle ($75-110^{\circ}$),
or tilted to an angle of $15^{\circ}$ between the beam and the sensor surface.

The data recorded at an impact angle of $15^{\circ}$ are also used for modeling 
charge drift and trapping in heavily irradiated sensors \cite{morris,vince}, 
and to measure the Lorentz angle \cite{andrei}
and the electric field within the sensors \cite{andrei2}.

\section{Signal Height}

The analogue information obtained from the readout chip was used to
study the signal height as a function of the sensor bias and the
irradiation fluence. To avoid  saturation of the electronics data were taken at
an angle of  $15^{\circ}$ between the beam and the sensor surface.
As the pitch is more than two times smaller than the sensor thickness
the collected charge per pixel is about 10000 electrons (most probable value) 
for an un-irradiated sensor.
The tilt of the sensor was such that the long clusters (``streets'')
run parallel to the pixel columns. The telescope information was used to select 
streets which run along the center of a column. 
With this selection charge sharing between
neighboring pixel columns was avoided. This also excludes
the two regions of reduced charge collection, the bias
dot and the metal line running along every second pixel column 
(see fig.~\ref{designs01}) from the analysis.
The charge of all pixels along the street was summed applying a
threshold of 2000 electrons. The charge distribution was fitted with
a Gaussian convoluted with a Landau. For each fluence and bias
voltage the most probable value was divided by the one obtained with
an un-irradiated sensor at 150\,V.

\begin{figure}
\centering\includegraphics[width=.8\linewidth]{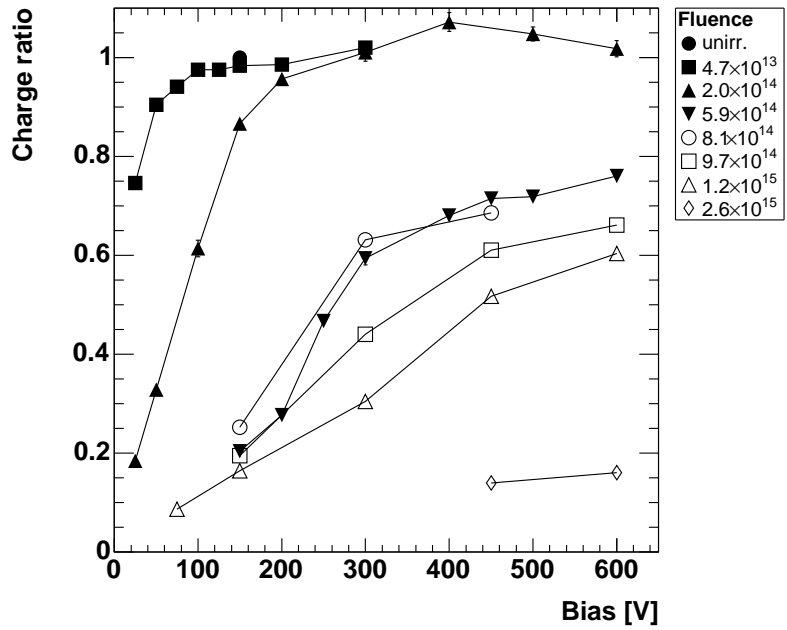}
\caption{Most probable signal as a function of the sensor bias. The
	signal of the un-irradiated sensor at $150\,$V is used as
	reference.
	\label{fig:sig_vs_V}}
\end{figure}

Figure~\ref{fig:sig_vs_V} shows this ratio as a function of the detector 
bias for several fluences.
The data were not corrected for possible differences in wafer thickness
or non-uniformities in the preamplifier gains which are estimated to
be at the few percent level. The increase of the ratio faster than
with the square root of the bias, typical for the ``n-in-n'' detectors after the
radiation induced space-charge sign-inversion (so called ``type inversion''),
is nicely visible. 
%At the bias usually referred to as ``full depletion''
%voltage the signal saturates.

The sensor irradiated to a fluence of $\Phi=2.6\times 10^{15}\,\Neq$
could only be operated up a maximum voltage of 600\,V at $-10^{\circ}\,$C.
At higher voltages noise exceeded 1000\,ENC and a reliable operation
was not possible.
% because it was not designed to accept
%the high leakage current (about 100\,nA per pixel).
%This could be solved either decreasing the feed back resistance of the
%preamplifier or cooling the sensor down to about  $-25^{\circ}\,$C.
%However, the first measure decreases the width of the analogue pulse
%to about 60\,ns which is our sampling time and no signal could be
%detected.
Therefore small data samples were recorded at 750\,V and 900\,V at
$-25^{\circ}\,$C to suppress the leakage current. The signal at
this voltages were slightly higher than at 600\,V.
%With the limited statistics it was not possible to perform a Landau
%fit to the puls height distribution. Average values in the order
%of $0.22-0.23$ compared to the un-irradiated sensor were reached
%which corresponds to about 5\,k electrons.

\begin{figure}
\centering\includegraphics[width=.8\linewidth]{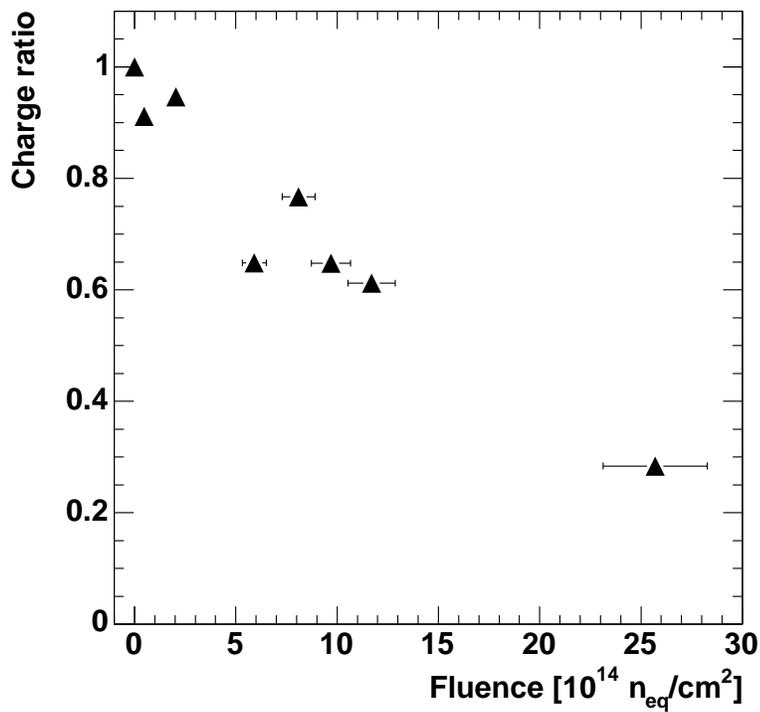}
\caption{Most probable signal as function of the irradiation fluence.
	\label{fig:sig_vs_Phi}}
\end{figure}

\begin{table}
\begin{center}
\begin{tabular}{cccccc}
\hline
$\Phi$&Bias&\multicolumn{2}{c}{Charge ratio}&\multicolumn{2}{c}{Efficiency}\\
$[10^{14}\,\Neq]$&[V]&[mip]&[ke$^{-}$]&0\,T&3\,T\\
\hline
0&150&1&22.4&$>0.999$&$>0.999$\\
0.47&100&0.91&20.3&$>0.999$&$>0.999$\\
2.0&200&0.95&21.3&$0.995$&$0.994$\\
5.9&400&0.65&14.6&$0.989$&$0.990$\\
8.0&450&0.77&17.2&$0.988$&\\
9.7&600&0.65&14.6&$0.990$&\\
11.7&600&0.61&13.7&$0.956$&$0.980$\\
25.7&600&0.28&6.3&$0.924$\protect\footnotemark&\\
\hline
\end{tabular}
\end{center}
\caption{Operation voltage and measured charge ratios
	for the sensors irradiated to different fluences. From this
	number the expected absolute charge is calculated.
	The detection efficiency is obtained using a pixel 
	threshold of 3\,k electrons.
	\label{tab:par}}
\end{table}
%
%\footnotetext{Using a pixel threshold of 2\,k electrons}%
%

Figure~\ref{fig:sig_vs_V} was used to
determine the best bias voltage for sensor operation. The 
spatial resolution very much depend on the charge sharing between 
neighboring channels caused by the Lorentz deflection in the 
magnetic field. As the Lorentz angle decreases with a higher 
bias \cite{andrei} the lowest bias voltage with a ``full'' signal 
collection was selected. The chosen voltages are listed in
Tab.~\ref{tab:par}. For those voltages large data samples were 
recorded with the beam perpendicular to the sensor surface.
The telescope prediction was used to select events in the pixel center
in order to avoid charge sharing and to exclude areas with reduced charge
collection. The charge distribution of the pixels predicted by the telescope 
was also fitted with
a Gaussian convoluted with a Landau and the most probable value
obtained from the fit was divided by the one obtained with
an un-irradiated sensor at 150\,V. The values of these charge
ratios are also listed in Tab.~\ref{tab:par}.

Figure~\ref{fig:sig_vs_Phi} shows the charge ratio as a function of
the fluence. In the fluence range relevant for CMS 
($0-12\times 10^{14}\,\Neq$) the sensor signal
will be above 12\,k electrons which is sufficient for a 
reliable operation at a threshold of 2500-3000\,electrons.
The readout chip forseen for CMS can probably be not operated at thresholds
much below this value for various reasons (occupancy of noise hits,
non-uniformities of the threshold adjustmet, time walk,
cross talk between digital and analogue parts etc.). Therefore
the signal of about 6\,k electrons delivered by the sensor irradiated 
to $2.6\times 10^{15}\,\Neq$ is probably too low for a reliable
and efficient operation,
especially if the signal charge is shared by two (or more) pixels.

\footnotetext{Using a pixel threshold of 2\,k electrons}%

The noise measured was about 400\,ENC for channels not connected to the
the sensor. This high noise was caused by the low setting of the feedback
resistors in the preamplifier and the shaper, needed to adjust the
dynamic range and the timing. Active pixels connected to the sensor
showed a noise of about 400\,ENC (un-irradiated) and 800\,ENC
($\Phi=1.2\times 10^{15}\,\Neq$ at 600\,V).

\section{Detection Efficiency}

In \cite{ieee03} it was shown that the bias dot and the
and, after irradiation, the region of the metal line
running along every second pixel column have a reduced charge
collection. Similar results have also been reported form 
other pixel detectors using a punch through dot \cite{Lari:2002wb}. 
Those regions, which
were excluded in the analysis shown in the previous section, degrade
the performance of the detector. There is a chance that a
particle that crosses the sensor in this region causes
a signal too small to exceed the threshold of a sparcified
readout.

To determine the effect of those regions on the sensor performance
data taken with a normal incidence angle were used. The beam telescope
is used to precisely predict the impact position on the sensor.
If the pixel predicted by the telescope or a direct neighbor
is above a certain threshold the track was counted as detected.
Due to dead time of the DAQ system the measured efficiency has
a systematic error which was estimated to about 0.1\,\%.
%Due to limited statistics and inefficiency caused by the DAQ
%system the determination of the inefficiency is only reliable
%for values above about 0.1\,\%.
%As the readout chip used in this test beam delivers non zero
%suppressed data of all pixels, the threshold can be applied in the
%off-line data analysis.

\begin{figure}[tp]
\centering\includegraphics[width=.8\linewidth]{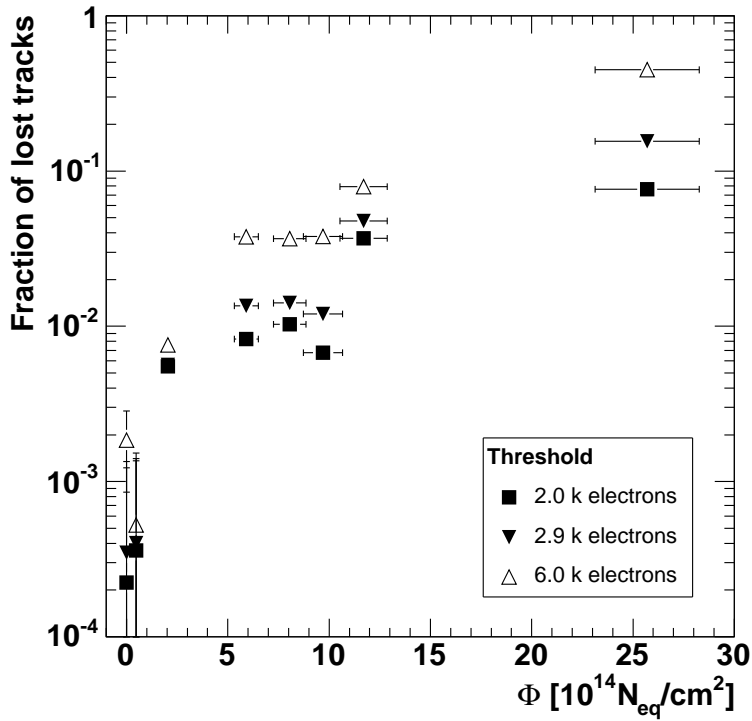}
\caption{Fraction of lost hits as function of the hadron
	fluence.
	\label{fig:eff_vs_Phi}}
\end{figure}

Figure~\ref{fig:eff_vs_Phi} shows the dependence of the detection inefficiency 
of the  sensors listed in tab.~\ref{tab:par} on the
radiation fluence for pixel thresholds of 2, 3 and 6\,k electrons without
magnetic field. Typically a threshold of 2-3\,k electrons is applied.
For fluences below $10^{15}\,\Neq$ and a threshold below 3\,k electrons the
fraction of lost tracks is well below 2\,\%, even with a threshold
of 6\,k electrons it does not exceed 5\,\%.

The signal of the sensor irradiated to $2.6\times 10^{15}\,\Neq$ has 
the most probable value of only about 6.2\,k electrons. For a reliable
operation of this sensor a threshold lower than half of this value is necessary.
With 2\,k electrons an efficiency of better than 90\,\% is reached. For
higher thresholds the efficiency decreased rapidly. The noise of this
detector was about $1000\,$e$^{-}$ which is too high for an operation
at such a low threshold. As the noise was to a large extend
caused by the inability of the readout electronics to accept the 
high leakage current, future measurements with sensors
irradiated to such high fluences will be carried out with modified readout chips
featuring an appropriate leakage current compensation.

\begin{figure}[tp]
\centering\includegraphics[width=1.\linewidth]{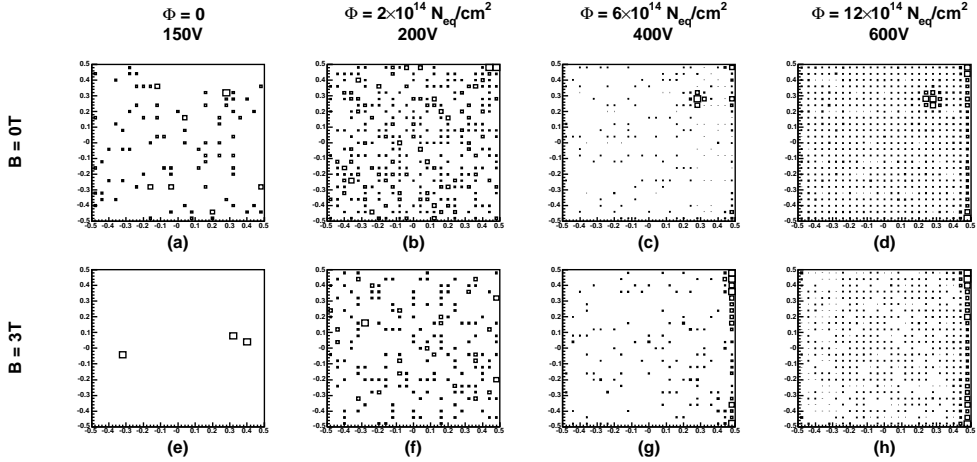}
\caption{Position of the tracks not detected by the pixel sensor
	for data recorded with (lower row) and without magnetic field 
	(upper row). Each
	plot represents the area of one pixel shown in fig.~\ref{designs01}.
	The bias dot is located in the top right corner, the aluminum line
	is placed along the right edge. The pixel
	threshold is set to 3\,k electrons.
	\label{fig:pix_pos}}
\end{figure}

The position of the lost hits within the pixel cell is shown
in fig.~\ref{fig:pix_pos} for different fluences and data with and 
without magnetic field. For the un-irradiated sensor and the
one irradiated to $2\times 10^{14}\,\Neq$ the hits below the
threshold of 3\,k electrons, are uniformly distributed over
the pixel area (see fig.~\ref{fig:pix_pos}a and b). 
This means that the charge collected in the less 
efficient regions is still safely above this threshold. If the
threshold is increased to about 6\,k electrons the undetected hits start
to be concentrated in the region of the bias dot.

For the devices with an irradiation fluence above a few $10^{14}\,\Neq$ 
the collected charge is significantly reduced by trapping. Additional
losses due to incomplete charge collection lead to an increased
inefficiency.
Hence the undetected hits are concentrated at the bias dot and
along the aluminum line (see fig.~\ref{fig:pix_pos}c and d). 
However, the total number of lost hits is small.

If a 3\,T magnetic field parallel to the horizontal axis
of the histograms in fig.~\ref{fig:pix_pos} is applied, the 
charge carriers are deflected by the
Lorentz force which is parallel to the vertical axis. This leads to
a distribution of the deposited charge along the vertical axis 
and reduces the influence of the small bias dot. Therefore the 
concentration of undetected hits around the bias dot is not present 
in fig.~\ref{fig:pix_pos}e--h. If a threshold of 6\,k electrons
is applied a slightly smeared ``image'' of the bias dot 
becomes visible in the highly irradiated sensors.
As the Lorentz drift of the signal charge is
parallel to the aluminum line, the number of undetected tracks in this
region is not effected by the magnetic field as shown in
fig.~\ref{fig:pix_pos}c, d, g and h. 
The total detection efficiency of the sensors
is not changed by the application of the magnetic field
(see tab.~\ref{tab:par}). It is still in a tolerable range below 5\,\%.

\section{Conclusions}

Silicon pixel sensors with \nit -side read out (\nit -in-\nit)
featuring moderated
\pit -spray  isolation have been irradiated up to proton fluences of
$2.6\times 10^{15}\,\Neq$.
The charge collection studies were performed with bump bonded samples
using a high energy pion beam. The total charge collected after
$1.2\times 10^{15}\,\Neq$ and a bias of 600\,V was about 60\,\%
compared to an un-irradiated sample.
After $2.6\times 10^{15}\,\Neq$ about 28\,\% of
the original signal could be collected. This result
is very encouraging with respect to possible upgrade scenarios for LHC.

The detection efficiency of the sensors is above 95\,\% after an
irradiation fluence of $1.2\times 10^{15}\,\Neq$ using a pixel threshold
of 3\,k electrons and a bias voltage of 600\,V. The bias dot and the
aluminum line connecting pixels originate the major source of inefficiency.
The influence of the dot is reduced if a magnetic field parallel to
the sensor surface is applied.

The tested sensors fulfill all requirements of the CMS experiment
and will be used in the barrel section of the pixel detector.

\section*{Acknowledgments}

The authors would like to thank Silvan Streuli from ETH Z\"urich
and Fredy Glaus from PSI for their enormous effort in bump bonding,
Kurt B\"osiger from the workshop of the University of Z\"urich for
the mechanical construction,
Maurice Glaser, Michael Moll, and Federico Ravotti from CERN for carrying
out the irradiation, Waclaw Karpinski from RWTH Aachen for providing the last
existing wafer of front-end chips, Gy\"orgy Bencze and Pascal Petiot from 
CERN for the H2-beam line support.
%Last but not least we gratefully acknowledge Roland Horisberger
%from PSI for explaining all details of the PSI30/AC30~readout chip. Without
%his advice this work would not have been possible.

\bibliographystyle{unsrt}
\bibliography{bib_rohe}

\end{document}